\documentclass[twocol]{ametsocV6.1}

\usepackage{amsmath,amsfonts,amssymb,bm}
\usepackage{mathptmx}
\usepackage{newtxtext}
\usepackage{newtxmath}
\usepackage{multirow}
\usepackage[utf8]{inputenc}

\title{Turbulence and energy dissipation from wave breaking}

    \authors{Jiarong Wu\aff{a,b},
    Stéphane Popinet\aff{c},
    Bertrand Chapron\aff{d},
    J. Thomas Farrar\aff{e},
    and Luc Deike\aff{a}\correspondingauthor{Luc Deike, ldeike@princeton.edu}}

    \affiliation{\aff{a}{Princeton University, Princeton, NJ, USA}\\
    \aff{b}{Courant Institute of Mathematical Sciences, New York University, New York, NY, USA} \\
    \aff{c}{Sorbonne Université, CNRS, Institut Jean Le Rond d’Alembert, F-75005 Paris, France} \\
    \aff{d}{Ifremer, LOPS, Brest, France} \\
    \aff{e}{Woods Hole Oceanographic Institution, Woods Hole, MA, USA} 
    }

\abstract{
Wave breaking is a critical process in the upper ocean: an energy sink for the surface wave field and a source for turbulence in the ocean surface boundary layer. We apply a novel multi-layer numerical solver resolving upper-ocean dynamics over scales from O(50cm) to O(1km), including a broad-banded wave field and wave breaking. The present numerical study isolates the effect of wave breaking and allows us to study the surface layer in wave-influenced and wave-breaking-dominated regimes. Following our previous work showing wave breaking statistics in agreement with field observations, we extend the analysis to underwater breaking-induced turbulence and related dissipation (in freely decaying conditions). We observe a rich field of vorticity resulting from the turbulence generation by breaking waves. We discuss the vertical profiles of dissipation rate which are compared with field observations, and propose an empirical universal shape function. Good agreement is found, further demonstrating that wave breaking can dominate turbulence generation in the near-surface layer. We examine the dissipation from different angles: the global dissipation of the wave field computed from the decaying wave field, the spectral dissipation from the fifth moment of breaking front distribution, and a turbulence dissipation estimated from the underwater strain rate tensor.  Finally, we consider how these different estimates can be understood as part of a coherent framework.} 

\begin{document}

\maketitle

%
%
%

%
\section{Introduction}

The top few meters of the oceanic boundary layer are heavily affected by the wind and surface waves. Wind is the major driving force for the large-scale ocean circulation \citep{ferrari_ocean_2009}. To a large extent, the wind stress acts on the upper ocean indirectly, by first amplifying the wave field, especially under high winds \citep{wang_wind_2004}. Then the locally saturated wave field undergoes wave breaking, which deposits momentum and energy into the current and upper-ocean turbulence \citep{sullivan_dynamics_2010}. Wave breaking thus acts as a sink for the wave energy but as a source for upper-ocean currents and turbulent kinetic energy (TKE). However, the energy pathway from the wind into the surface waves and eventually into the upper ocean-current (mean flow) and turbulence (fluctuation) is not well characterized, largely due to the difficulty of making measurements in the vicinity of the sometimes violently moving air-sea interface. 

This paper is devoted to understanding the process of turbulence generation and dissipation associated with broad-banded surface-wave breaking in the ocean surface layer, i.e. the uppermost few meters of the ocean boundary layer (OBL). Non-breaking waves can interact with the upper ocean through mechanisms such as Stokes drift and Langmuir circulation \citep{sullivan_dynamics_2010}, but the focus of this paper is on breaking-wave-induced turbulence.

Many previous works have focused on the vertical profile of energy dissipation rate under different wind forcing and wave conditions. Despite various technical challenges, field observations of dissipation rate profiles in the OBL surface layer have been conducted over the years \citep[][etc.]{anis_surface_1995, drennan_oceanic_1996, terray_estimates_1996, soloviev_observation_2003, gerbi_observations_2009, esters_turbulence_2018, zippel_parsing_2022, miller_scaling_2023}. In these observations, turbulence is measured with either fixed sensors or rising profilers, and the dissipation rate is extracted using methods based on turbulence spectra or structure functions. These measurements suggest that wave breaking enhances dissipation up to an order of magnitude compared to law-of-wall prediction. Most measurements also suggest a deviation from the law-of-the-wall scaling ($z^{-1}$) in an intermediate range of depth \citep[][and thereafter T96]{terray_estimates_1996}, although contradictory evidence exists \citep{esters_turbulence_2018}. There are fewer reports or estimates of breaking-induced current \citep[except for][]{kudryavtsev_vertical_2008}, likely due to the difficulty of isolating the breaking-induced component of the current, as well as the inherent difficulty of measuring currents in the presence of the large wave orbital currents.

Since the effects of wave breaking on upper-ocean current and turbulence is not yet fully understood, their inclusion in modeling of the OBL is not routine and is limited to low-order representations. In a Reynolds-averaged Navier-Stokes (RANS) type of framework, one typically models breaking as a source term in the turbulent kinetic energy (TKE) equation and/or momentum equation, with the goal of reproducing observed dissipation profiles. The input can be either at the surface for TKE \citep{craig_modeling_1994} or with a prescribed injection depth \citep{kudryavtsev_vertical_2008, sullivan_surface_2007}, for both TKE and momentum. The other modeling choice involves how to represent the spectral distribution of breaking waves; either waves of all scales are considered together or some empirical breaking distribution needs to be prescribed. A nice summary of different modeling approaches is given in \cite{rascle_note_2013} and the effects of entrainment profile and breaker distribution on dissipation profile are discussed.

More sophisticated large eddy simulation (LES) modeling \citep[e.g.,][]{sullivan_surface_2007} is of higher resolution and fidelity than RANS models, but it is not routinely used in practice due to the higher computational cost. In \cite{sullivan_surface_2007}, breakers are randomly sampled from an empirical breaking probability distribution with certain global constraints of momentum and energy conservation, and each breaker is modeled by a self-similar momentum and energy entrainment profile. In addition, wave-current interaction is modeled by including a vortex force. This work suggests that overall OBLs behave differently from purely shear-driven turbulent boundary layers due to the existence of surface waves, and the effects of realistic intermittent wave breaking depends on the breaking distribution. 

The current study aims to numerically investigate the breaking distribution and the associated underwater turbulent field as an emergent phenomenon of broadband wave spectra. In our previous paper \citep[][hereafter W23]{wu_breaking_2023} we introduced the use of the novel multi-layer numerical method \citep{popinet_vertically-lagrangian_2020} for modeling breaking-wave fields. The simulations have similar scale and resolution to \cite{sullivan_surface_2007}, but a fundamentally different treatment of wave breaking. In our framework, the breaking events are not sampled from a prescribed distribution but emerge naturally through focusing and are detected using a geometric criterion of the free surface. No subgrid entrainment models are needed since we are directly approximating the Navier-Stokes equations and are resolving the breaking events and their effects. In W23, we characterized the breaking kinematics and statistics based on free surface elevation following the framework first introduced by \cite{phillips_spectral_1985} and found good agreement with field observations \citep[][etc.]{schwendeman_wave_2014, sutherland_field_2013, kleiss_observations_2010}. Based on the numerical results, we proposed a new scaling for the distribution of breaking crest length using wave-spectrum-related quantities (peak phase speed and mean square slope). By showing the predictive power of the proposed scaling on existing observational data, we further confirmed the direct link between the surface wave spectrum and breaking distribution.

In this paper, we investigate the upper-ocean turbulent fields associated with broadbanded wave breaking. We first show vorticity generation by breaking (concentrated in the top few meters of wave-influenced layer). Then we focus on an analysis of the vertical structure of the dissipation rate (estimated from the strain rate tensor) and the total dissipation of the system. The effect of wave breaking is isolated due to the fact that we are analyzing short time scale behaviors (O(10) peak wave periods) without a development of wave-current interaction and that by construction there is no wind forcing. Results are compared with field observations of dissipation rate profiles and discussed in the context of OBL modeling.

\section{Numerical Methods}
\subsection{Governing equation and numerical solver}
\begin{figure}
    \centering
    \includegraphics[width=1\linewidth]{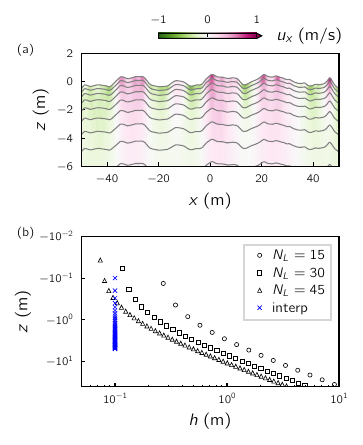}
    \caption{(a) Illustration of the layered discretization for $N_L=15$ (for only part of the domain). Color shows the horizontal velocity and the gray lines show the layers. (b) The geometric distribution of layer thickness. The blue marks are the uniformly-spaced interpolation points used in analysis in Figure \ref{fig:aver_layer_interp}.}
    \label{fig:coord}
\end{figure}

The numerical setup follows W23 and uses the multi-layer numerical solver \citep{popinet_vertically-lagrangian_2020,wu_ocean_2023}. An example of the multi-layer discretization with the free surface is shown in Figure \ref{fig:coord}(a).  We solve the Navier-Stokes equations written as a set of layered equations \citep{popinet_vertically-lagrangian_2020}, with the most general form of the governing equations being given by:
\begin{align}
\begin{split}\label{eqn:ml1}
\partial_t h_l + \nabla_H \cdot (h\boldsymbol{u})_l ={} & 0
\end{split}\\
\begin{split}\label{eqn:ml2}
\partial_t (h\boldsymbol{u})_l + \nabla_H \cdot (h\boldsymbol{u}\boldsymbol{u})_l = {} & -g h_l \nabla_H \eta - \nabla_H(h\phi)_l + \\
[\phi\nabla_H z ]_l + & [\nu_1 \partial_z \boldsymbol{u}]_l + \nu_2 \nabla^2_{H} \boldsymbol{u}
\end{split}\\
\begin{split}\label{eqn:ml3}
\partial_t (hw)_l + \nabla_H \cdot(hw \boldsymbol{u})_l ={}& -[\phi]_l + [\nu_1\partial_z w]_l +  \nu_2 \nabla ^2_{H} w
\end{split}\\
\begin{split}\label{eqn:ml4}
\nabla_H \cdot (h\boldsymbol{u})_l + [w-\boldsymbol{u}\cdot \nabla_H z ]_l ={}& 0
\end{split}
\end{align}
with $l$ the index of the layer, $h$ its thickness, $z$ its vertical position, $\boldsymbol{u}=(u,v)$, $w$ the horizontal and vertical components of the velocity, $\phi$ the non-hydrostatic pressure (divided by density),  $g = 9.8 \; \mathrm{m/s}$ the gravitational acceleration, $\nu_1$ and $\nu_2$ the vertical and horizontal (kinematic) viscosity coefficients. The surface elevation $\eta = z_b + \sum_{l=0}^{N_L} h_l$, and the $[\;]_l$ operator denotes the vertical difference, i.e. $[f]_l = f_{l+1/2} - f_{l-1/2}$. Equation \ref{eqn:ml1} means that the layer thicknesses $h_l$ follow material surfaces (i.e. the discretization is vertically Lagrangian). Equations \ref{eqn:ml2} and \ref{eqn:ml3} are the horizontal and vertical momentum equations. Equation \ref{eqn:ml4} is the mass conservation equation (used to implement the incompressibility conditions). 
The anisotropy of the grid makes it a reasonable option to use different horizontal and vertical viscosities. We specify a small value of physical viscosity while the dissipative effects in the numerical system are dominated by numerical viscosity and the gradient limiter (equation \ref{eqn:breaking_model}).


In terms of numerical resolution, the domain size is ($L_x, L_y, L_z$) = (200, 200, 40) m with $(N_H,N_H,N_L)$ grid points. The horizontal grid is evenly spaced and we use $N_H=1024$ (for a numerical convergence test on horizontal resolution see W23). The vertical grid spacing follows a geometric series and the averaged layer depths for $N_L=15,30,45$ are shown in Figure \ref{fig:coord}(b) (the thickness of the top-most layer is 0.27, 0.12, 0.07 m, respectively). The horizontal resolution is about five times finer than in the LES of \cite{sullivan_surface_2007}. The vertical layers are more concentrated towards the surface \citep[thickness ratio 1.2 between adjacent layers instead of 1.012 in][]{sullivan_surface_2007}, since we are directly resolving the waves. Aside from the breaking model described in the next section, there is no explicit subgrid-scale turbulence model: the effect of any unresolved subgrid-scale structure is assumed to be adequately approximated by numerical diffusion (see W23 for a discussion).

\subsection{Modeling of wave breaking}
The dissipative effect on waves due to wave breaking is modeled by a gradient limiter. The horizontal gradient $\partial z/\partial x$ for any layer in equation \ref{eqn:ml2} and \ref{eqn:ml4} is limited by a maximum value $s_\text{max}$:
\begin{equation}
  \partial z/ \partial x = \left\{
    \begin{array}{ll}
      \partial z/\partial x, & |\partial z/\partial x| \leq s_\text{max}  \\[2pt]
      \text{sign}(\partial z/\partial x)s_\text{max}, & |\partial z/\partial x| > s_\text{max}.
  \end{array} \right. \label{eqn:breaking_model}
\end{equation}
Note that the wave slope (or the slope between any layer) itself is not affected by this limiter (only its gradient used in equations \ref{eqn:ml2} and \ref{eqn:ml4}). The solutions are not particularly sensitive to the value of $s_{\mathrm{max}}$ and a value of 0.577 is chosen (see W23 for validations). There can be no overturning of the free-surface in the multi-layer model as surface elevation $\eta$ is a single-valued function, and the breaking dynamics is thus shock-like (but with additional dissipative and dispersive effects taken into account compared to a pure shallow-water model). 

\subsection{Broadband wave field simulations}

We initialize the wave field with a directional wavenumber spectrum $\phi(k,\theta)$ derived from an azimuth-integrated wavenumber spectrum $\phi(k)$ typical of wind waves \citep[see e.g. observations from][]{lenain_measurements_2017}.
\begin{equation}\label{eqn:spectrum_init}
\phi(k) = Pg^{-1/2}k^{-2.5}\exp[-1.25(k_p/k)^2].
\end{equation}
The directional spreading is proportional to $\cos^N(\theta)$ with $N=5$. The value of $P$ (of dimension ms$^{-1}$) controls how energetic the wave field is, while $k_p$ is the peak wavenumber of the spectrum. We use $k_p=2\pi/40 = 0.157 \;\mathrm{m^{-1}}$ for all the cases presented here. For this peak wavenumber, the computational domain is large enough both in the horizontal direction to avoid confinement effects, and in the vertical direction to ensure that the wave dynamics do not depend on the bottom boundary.

\begin{table}[!h]
    \centering
    \begin{tabular}{|c|c|c|c|c|}
        Label & $P \; \mathrm{(ms^{-1})}$ & $k_p\; \mathrm{(m^{-1})} $ & $\sigma$ & $H_s \; \mathrm{(m)}$\\ 
        P008 &  0.08 & \multirow{5}{*}{0.157}& [0.099, 0.103] & [0.983, 0.986] \\ 
        P01 &  0.1 & & [0.116, 0.122] & [1.097, 1.104] \\ 
        P016 &  0.16 & & [0.145, 0.147] & [1.341, 1.359] \\ 
        P02 &  0.2 & & [0.150, 0.154] & [1.442, 1.460] \\ 
        P03 &  0.3 & & [0.161, 0.165] & [1.697, 1.735]
        \end{tabular}
    \caption{A compilation of cases and their steepness parameters. The range of $\sigma$ and $H_s$ indicate the variability within the sampling window.}
    \label{tab:cases}
\end{table}

\begin{figure*}[h]
    \centering
    \includegraphics[width=0.9\linewidth]{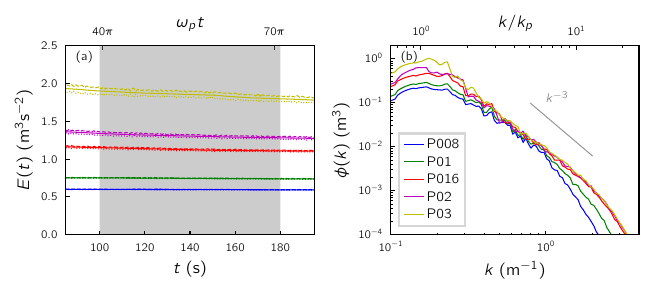}
    \caption{(a) Energy evolution of each case showing the system is slowly decaying without external forcing. Because equipartition of kinetic and potential energy is expected for linear waves, we show twice the kinetic energy, $2E_k$ (dashed lines); twice the potential energy, $2E_p$ (dotted lines); and the total energy, $E = E_k + E_p$ (solid lines). The sampling period is indicated with the gray shading. (b) The spectra of each case at the start of the sampling period $t=100$ s. For both plots, the color of lines indicate different initial spectrum $P$ value listed in Table \ref{tab:cases}. The bottom horizontal axis shows dimensional variables while the top horizontal axis shows non-dimensional variables normalized by peak wavenumber $k_p$ or peak frequency $\omega_p$.}
    \label{fig:energy}  
\end{figure*}

The initial wave field is a superposition of linear waves $\eta = \sum_{i,j} a_{ij}\text{cos}(\psi_{ij})$, with the amplitude $a_{ij} = [2\phi(k_{xi},k_{yj})dk_xdk_y]^{1/2}$ and the initial random phase $\psi_{ij} = k_{xi} x + k_{yj} y + \psi_{\text{rand}\:ij}$. 
The corresponding orbital velocity is initialized similarly according to the linear wave relation. Only a few long-wave components are used for initialization. After initialization, the wave field evolves, undergoes intermittent breaking due to random focusing, and develops the short-wave tail over time. The evolution of the wave field and the spectral shape are discussed in detail in W23. Here we focus on the later stage when the spectral shape is quasi-stationary while the energy of the wave field slowly decays. 

The value of $P$ roughly models different wind speeds, assuming similarity in the shape of the spectra. It is chosen to range from 0.08 (almost no breaking) to 0.3 (steep actively breaking) in the simulations, and is used for labeling cases listed in Table \ref{tab:cases}. 
We also define two diagnostic variables, the root-mean-square-slope (rmss)
\begin{equation}
    \sigma = (\int_0^{\infty} k^2 \phi(k)dk)^{1/2},
\end{equation}
and the significant wave height 
\begin{equation}
    H_s = 4\langle \eta^2 \rangle^{1/2} = 4(\int_0^{\infty} \phi(k)dk)^{1/2}.
\end{equation}
where the angle brackets denote horizontal averaging. In practice, the integrals are taken over numerically resolved wavenumbers. Table \ref{tab:cases} summarizes the values of $\sigma$ and $H_s$ for different initial values of $P$. In general, the initial condition $P$, rmss $\sigma$, and significant wave height $H_s$ are positively correlated, but $\sigma$ and $H_s$ characterize different aspects of the wave field. The $H_s$ value indicates the energy content, and is largely determined by the low wavenumber spectral peak, while the $\sigma$ value indicates the roughness of the surface, and is heavily influenced by the highwavenumber part of the spectrum. Their values for a given initial $P$ are specific to the current setup. 

Figure \ref{fig:energy}(a) shows the energy evolution for each case with the sampling window indicated in gray shade. Figure \ref{fig:energy}(b) shows the spectral shape at the start of the sampling window (t=100). Here, the energy of the system $E$ is defined as the sum of volume-integrated potential energy ($E_p = \iint g\eta^2 dxdy/2$, dashed lines) and kinetic energy ($E_k = \iiint |\mathbf{u}|^2 dxdydz$, dotted lines). The kinetic energy is slightly higher than the potential energy because of the existence of non-wave motion. The wave energy is $E_w = 2E_p = \iint g\eta^2 dxdy$ assuming equipartition of potential and kinetic energy for the waves. The energy of the wave field steadily decays without any external forcing, and we see the fastest decay for the steepest case due to more frequent breaking. The small discrepancy between the total energy of the system $E$ and the wave energy $E_w$ does not affect the discussion of the decay rate, as we found $dE/dt \approx dE_w/dt$. 

The total integration window in the quasi-stationary phase is relatively short (on the order of 15 peak wave periods). There are two main reasons for this choice: first, in this study we focus on the short-term effects of breaking in generating current and turbulence; second, we are limited by the dissipative nature of the simulation (which is an intrinsic limitation of the current setup). Future implementations will add boundary wind-like forcing to balance the dissipation so that the long-term wave effects on the upper-ocean boundary layer can be studied. The range of rmss and significant wave height in the sampling window are listed in Table \ref{tab:cases}.

\begin{figure}[ht]
    \centering
    \includegraphics[width=1\linewidth]{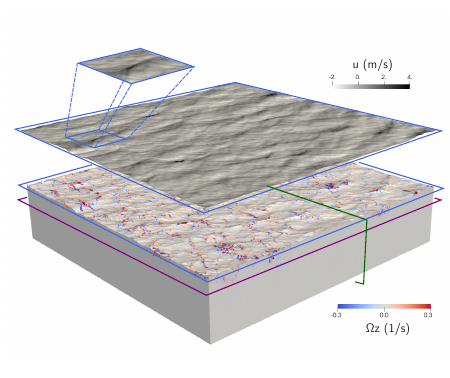}
    \caption{A 3D illustration of the simulation. The surface velocity field $u$ is shown in gray scale and offset by a distance from the surface vorticity field. The zoomed-in view shows one of the detached breaking fronts based on Gaussian curvature (see W23 for detail). The vorticity component $\Omega_z$ is shown in Blue-Red color. The blue frame indicates the surface; the purple frame indicates depth $z\approx - 3H_s$; the green framed indicate the $x-z$ plane. The same color notation for plane orientation is used in Figure \ref{fig:vorticity}.}
    \label{fig:3D}  
\end{figure}

\section{Diagnosis}
Figure \ref{fig:3D} shows a 3D visualization of the simulation that demonstrates the surface geometry of the broadband wave field and the rich vorticity field. In the following sections, we describe the diagnosis of both the breaking statistics at the surface and the current and turbulence in the interior.

\subsection{Breaking front distribution}
Despite there being no overturning in the simulation, the breaking fronts are characterized by the sharp ridge-like features of the surface. In the rest of the paper, we refer to this localized sharpening of the surface in the numerical simulation as ``breaking". Such sharpening induces strong dissipation and generates vorticity similar to actual breaking events. We perform breaking front detection based on a Gaussian curvature criterion, and the procedure and sensitivity studies are discussed in detail in W23. 

\begin{figure*}[ht]
    \centering
    \includegraphics[width=0.9\linewidth]{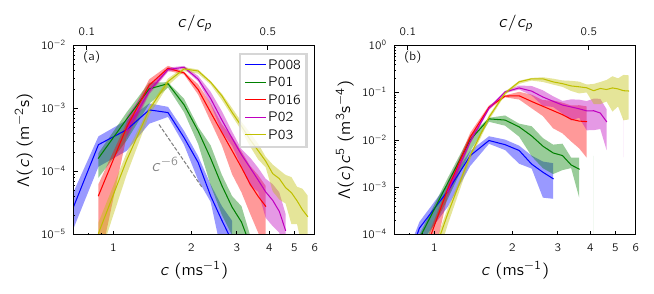}
    \caption{(a) Breaking distribution $\Lambda(c)$ for increasing wave intensity. (b) Pre-multiplied breaking distribution $c^5\Lambda(c)$. For both plots, the bottom horizontal axis shows dimensional variables while the top horizontal axis shows non-dimensional variables normalized by peak phase speed $c_p$.}
    \label{fig:lambdac}  
\end{figure*}

The main result of W23 is that the distribution $\Lambda(c)$, defined as the length of breaking front per unit area moving with speed in the vicinity of $c$ \citep{phillips_spectral_1985}, are self-similar for wave fields of similar spectral shape but different steepness $\sigma$ and peak phase speed $c_p$. A scaling using only wave field features (i.e. peak wave phase speed $c_p$ and rmss $\sigma$) was proposed and tested on field observations. Figure \ref{fig:lambdac} shows the $\Lambda(c)$ distribution, as well as pre-multiplied by $c^{5}$, without re-scaling. The fifth moment of $\Lambda(c)$ is theoretically related to wave energy dissipation, which will be discussed in section \ref{sec:dissipation}. Finite sample size introduces uncertainty in the statistics, especially for less frequent breakers at large $c$. We estimate the uncertainty in both $\Lambda(c)$ and $\Lambda(c)c^5$ by further dividing the sample window into four chunks of the same size and compute the maximum and minimum among the four sub-windows. The range is indicated by the shading in Figure \ref{fig:lambdac}. More on breaking front distribution can be found in W23 while here we focus on the underwater dynamics, including current and turbulence generation, as well as energy dissipation.

\subsection{Characterizing turbulence through vorticity}

\begin{figure}[h]
    \centering \includegraphics[width=1\linewidth]{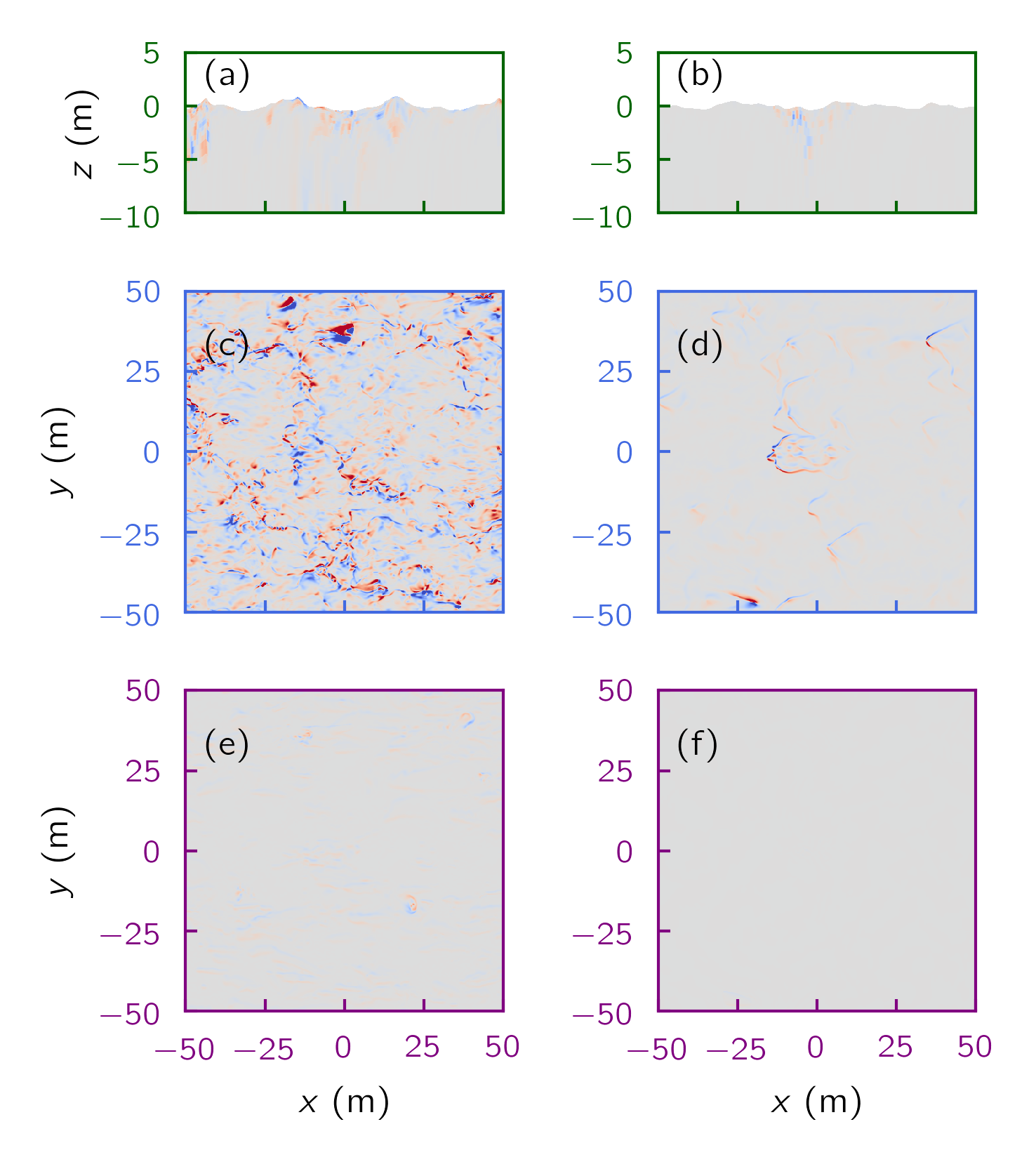}
    \caption{(a) An example of a slice of $x-z$ plane at $y=0$ for a moderately steep case P02 at $t=120$. (c) The surface layer. (e) The layer of average depth around $3H_s$. (b,d,f) Same as (a,c,e) but for the least steep case P008.}
    \label{fig:vorticity}
\end{figure}

The surface layer of the OBL consists of both wave and turbulent motions. It is a real challenge to separate them in field observations, which is another reason why numerical simulations can be valuable. The wave orbital motion is largely irrotational, while the turbulence is characterized by high vorticity. Therefore, we use vorticity as an indicator for turbulence intensity. The vorticity vector $\boldsymbol{\Omega}=(\Omega_x, \Omega_y, \Omega_z)$ is defined as
\begin{equation}
   \boldsymbol{\Omega} = (\partial_y w - \partial_z v, \partial_z u - \partial_x w, \partial_x v - \partial_y u).
\end{equation}
We use $\boldsymbol{\Omega}$ to denote vorticity to distinguish it from the wave angular frequency $\omega$.

As shown by the 3D visualization in Figure \ref{fig:3D} and the 2D slices in Figure \ref{fig:vorticity}, intense patches of vorticity can be found near the surface, with a fast decay at larger depths. At $z\approx -3H_s$ (shown by the purple framed subplots), only a very weak signature of paired vortices can be seen. Figure \ref{fig:vorticity} also provides a comparison between the least steep (P008) and the moderately steep cases (P02) that we simulated. The least steep case with very infrequent breaking shows much less turbulence generation and shallower penetration compared to the moderately steep case. This demonstrates that the vorticity is indeed generated by breaking related processes. In the steep cases, intense vorticity generation is concentrated under the breaking crests, but seems to be transported further downwards by orbital wave motion.



\subsection{Dissipation rate}
Another quantity of interest is the dissipation rate in the surface layer. In particular, we are interested in the vertical profile of dissipation rate, and how it varies for wave fields of different steepness and thus breaking frequency and intensity. In order to diagnose the dissipation rate, we compute the strain rate tensor $\mathbf{S}$, whose components are
\begin{equation}
s_{ij} = \frac{1}{2}(\frac{\partial{u_i}}{\partial{x_j}} + \frac{\partial{u_j}}{\partial{x_i}}),
\end{equation}
where $u_1=u, u_2=v, u_3=w$ and $x_1=x, x_2=y, x_3=z$. The dissipation rate (per unit volume) is defined as 
\begin{equation} \label{eqn:epsilon}
    \epsilon = 2 \nu s_{ij}s_{ij}.
\end{equation}
where $\nu$ is the viscosity and the Einstein summation convention is used. 

\begin{figure}[h]
    \centering
    \includegraphics[width=1\linewidth]{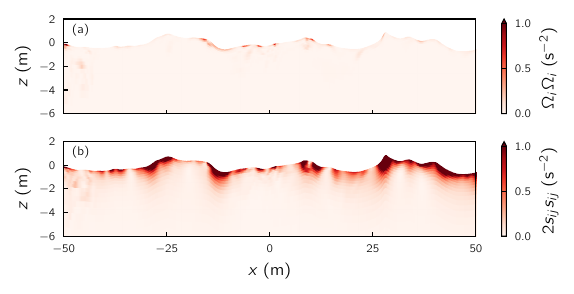}
    \caption{Heatmap of (a) square of vorticity and (b) two times square of strain rate tensor to identify areas of high dissipation.}
    \label{fig:Omega_vs_sij}
\end{figure}

In certain flow regimes, for example homogeneous high Reynolds number turbulence, the square of vorticity is used as a proxy for the dissipation rate in a globally averaged sense \citep{tennekes_first_1972, donzis_dissipation_2008}. In other words, $\overline {\Omega_i \Omega_i} \approx \overline {2 s_{ij}s_{ij}} $ where the overline denote spatial averaging. However, we found that this relation does not apply to the flow simulated here. Figure \ref{fig:Omega_vs_sij} shows both the square of vorticity $\Omega_i \Omega_i$ and two times the square of the strain rate tensor $2s_{ij} s_{ij}$. The hot spots of $\Omega_i \Omega_i$ are localized and indicate the turbulent region, while the high strain rate area also includes the near-surface region.  In other words, in addition to the turbulence that contributes to the dissipation, there is an area of high strain rate near the free surface due to wave motion, which is an order of magnitude larger than the square of vorticity.


\subsection{Note on choice of vertical coordinate}
Due to the difficulties of taking measurements near the moving sea surface, sampling of near-surface turbulence typically needs to be made in a wave-following coordinate system. The dissipation rate measurement and its interpretation might differ based on whether it is made in a wave-following or absolute vertical coordinate \citep[see e.g.][]{soloviev_near-surface_2014}. 

\begin{figure*}[!ht]
    \centering
    \includegraphics[width=0.9\linewidth]{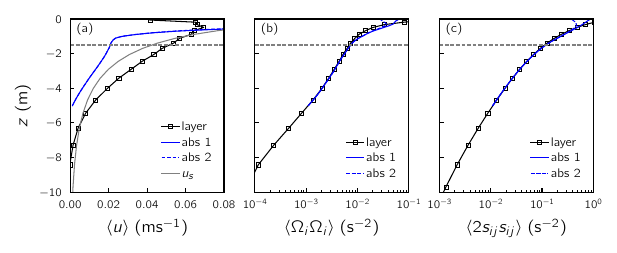}
    \caption{Vertical profiles of horizontally averaged quantities $\langle \cdot \rangle$ in layer coordinates (squares) v.s. absolute coordinates (blue lines). For the interpolated absolute coordinate, abs 1 denotes averaging in only the subsurface area while abs 2 denotes averaging over the whole horizontal area. (a) Velocity component $u$. Gray solid line shows the Stokes drift computed according to equation \ref{eqn:stokes}. (b) Vorticity magnitude squared. (c) Two times strain rate tensor magnitude squared. All plots here are for layer number $N_L=30$ but cases of $N_L=15$ and 45 show similar behaviors.}
    \label{fig:aver_layer_interp}
\end{figure*}

We use our simulations as a way to test the effects of vertical sampling choices on mean statistics. To do so, we compute horizontally averaged quantities of interest, denoted with angled brackets $\langle \cdot \rangle$, using both kinds of vertical coordinates (layer or surface-following and an absolute or Eulerian vertical coordinate) and plot their vertical profiles in Figure \ref{fig:aver_layer_interp}. In the ``layer coordinate'' (denoted by black lines and markers), we simply average the quantity in each layer $l$, and use the layer averaged vertical position $\overline{z_l}$ to draw the 1D profile. This can be considered as approximately wave-following. In the ``absolute coordinate'' (denoted by blue lines), we interpolate the field onto a regular Cartesian coordinate and then average horizontally. Since the horizontal grid spacing is constant, this interpolation is 1D in $z$. The horizontal averaging needs further clarification for depths that are not always subsurface (above the gray dashed line around $-1.5$ m), taken either only for the water phase (abs 1) or for the whole horizontal plane (abs 2). There is no such ambiguity for the layer coordinate.

Interestingly, the coordinate choice makes a significant difference for analyzing the horizontal velocity profile (Figure \ref{fig:aver_layer_interp}a), but not for the vorticity and strain rate tensor (Figure \ref{fig:aver_layer_interp}b,c). 
This is because when we compute the mean of velocity $u$ in a layer coordinate, we are sampling relatively higher magnitude positive regions and lower magnitude negative regions (an effect similar to the origin of Stokes drift). The vertical profile thus becomes a hybrid of Lagrangian and Eulerian mean. This effect is well known in current meters attached to surface-following buoys \citep{pollard_interpretation_1973} and is expected to cause a difference equal to half the Stokes drift at the surface, decaying with depth more gradually than the Stokes drift profile, which is qualitatively what we see in Figure \ref{fig:aver_layer_interp}(a). On the other hand, there is not such correlation between the surface elevation and the turbulence statistics, so the mean of vorticity and strain rate are not affected. We thus use layer averaging in Figure \ref{fig:diss-scaling}.

To make an order-of-magnitude comparison, we also plot the Stokes drift $u_s$ in Figure \ref{fig:aver_layer_interp}(a), estimated using \cite{kenyon_stokes_1969}
\begin{equation} \label{eqn:stokes}
    u_s(z) = 2g^{1/2} \int k^{3/2}\phi(k) e^{2kz} dk, 
\end{equation}
where $\phi(k)$ is the uni-directional spectrum. It is worth pointing out that even the strictly Eulerian mean drift current (blue line) is comparable to the Stokes drift, despite being smaller in magnitude and of different vertical structure. This closeness in magnitude might have implications for the estimation of currents on a large scale, which to our knowledge has not been widely studied. The current profile in the upper ocean can of course be generated through various mechanisms (tide, submesoscale instability, etc.), but here we are isolating the effects of wave breaking. 

Finally, for the immediate surface regions that are only partially subsurface (above -1.5 m), there is a small discrepancy between the three different ways of computing the horizontal mean very close to the surface (above -0.5 m). The ``abs 1'' method is the most consistent with flat-surface simulations, since it makes sure that the volume-averaged quantities are equal, while the layer method might be closer to the practices in field measurements. 


\section{Turbulent boundary layer mean statistics}
In the following section, we characterize the underwater breaking-modulated turbulent boundary layer. Again, we focus on the quasi-stationary (developed) stage. First, we comment on the velocity spectrum observed in the simulation. Then we discuss the scaling of vertical dissipation rate profiles, and compare with field observations. We note that even for the developed stage, the time scale considered here is too short for the development of Langmuir turbulence (on the order of $40$ wave periods), especially with the lack of a pre-existing background current. We thus interpret the results as being representative of conditions dominated by breaking instead of by wave-mean flow interaction or Langmuir turbulence.

%
 

\subsection{Turbulence spectrum}
An advantage of our simulation is that we can directly examine the spatial structure of breaking-generated turbulence, partially represented by the horizontal wavenumber spectra. Observational studies often use turbulence spectra (or, similarly, structure functions) to infer dissipation rate \citep[][etc.]{gemmrich_near-surface_2004, thomson_wave_2012, sutherland_field_2015}, based on theories of the inertial range of isotropic turbulence. However, they are often limited to single-point measurement in time, and, in some cases, the wavenumber spectrum is estimated from the frequency spectrum using Taylor's frozen-field hypothesis. There is also the issue of sampling and measurement uncertainty due to the intermittent nature of turbulent dissipation \citep{soloviev_observation_2003, sutherland_field_2015} and flow distortion and turbulence generation by the instrument itself \citep[e.g.,][]{zippel_moored_2021}. 

\begin{figure}[!h]
    \centering
    \includegraphics[width=0.8\linewidth]{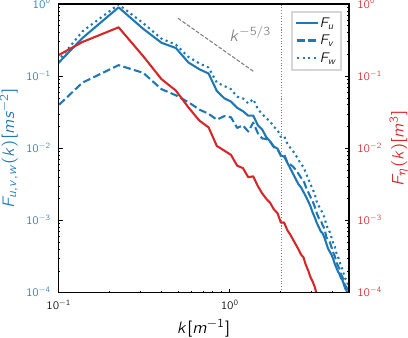}
    \caption{Spectrum of three velocity components (blue) and surface height spectrum (red). Plotted at surface layer z = -0.058 m ($N_L=30$) during the developed stage. The gray dotted line is the numerical resolution limit calibrated in W23. Gray dashed line indicate the $k^{-5/3}$ power law.}
    \label{fig:spectrum1}
\end{figure}

Since we have the full 3D velocity field, no conversion from frequency to wavenumber spectrum is needed. The 2D spectrum $F(k_x,k_y)$ is calculated in each layer and then azimuthally averaged to get $F(k)$. Only the top layer (at average depth $-0.058$ m) result is shown in Figure \ref{fig:spectrum1}, where blue curves are velocity spectra and the red curve is the surface wave energy spectrum for comparison.

The velocity spectrum is dominated by wave motion at low wavenumber, again highlighting the challenge of separating wave and turbulence motion \citep[e.g.,][]{gerbi_observations_2009}. At higher wavenumber, the velocity spectrum is less steep than the wave spectrum, indicating the presence of small-scale turbulence. Since the turbulence is solely generated by breaking waves, it is clearly anisotropic - fluctuations of the $v$ component (perpendicular to the wave traveling direction) contain lower energy. Spectra of $u$ and $w$ exhibit slopes close to that of an isotropic turbulence spectrum with a $k^{-5/3}$ power law. However, it is in a small range of high wavenumber limited by numerical resolution. Overall the flow shows strong anisotropy, which brings into question the validity of spectrum-derived estimates of dissipation $\epsilon$ that is driven by wave breaking. Therefore, we proceed with using directly computed velocity gradient information (strain rate tensor) for dissipation rate estimation.

 

\subsection{Vertical profile of dissipation and comparison with observations} \label{sec:diss-profile}
\begin{figure*}
    \centering  
    \includegraphics[width=0.9\linewidth]{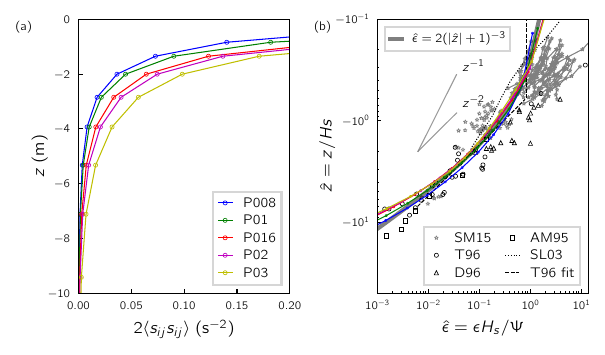}
    \caption{(a) Layer-averaged squared strain rate tensor profiles in linear-linear scale for different cases. (b) Dissipation rate profile in log-log scale normalized by $H_s$ and $\Psi$. Plotted field observations: \cite{anis_surface_1995} (AM95), \cite{terray_estimates_1996} (T96), \cite{drennan_oceanic_1996} (D96), \cite{soloviev_observation_2003} (SL03), \cite{sutherland_field_2015,sutherland_measuring_2015} (S15). The T96 fit is $\hat{\epsilon} = 0.3\hat{z}^{-2}$ below z = -0.6 $H_s$ and constant above.}
    \label{fig:diss-scaling}
\end{figure*}

Vertical profiles of dissipation rate are often used to compare field measurements with various OBL models \citep{rascle_note_2013}. The squared strain rate tensor $\langle s_{ij}s_{ij}\rangle$ provides a direct estimate of dissipation (Equation \ref{eqn:epsilon}). In Figure \ref{fig:diss-scaling}(a) we plot the vertical profiles of layer-averaged squared strain rate tensor for different cases. With steeper wave fields and more frequent breaking (darker orange lines), there is stronger breaking-induced turbulence that extends to greater depths. The rapid decay with depth is highlighted by the linear scale.


In Figure \ref{fig:diss-scaling}(b), the profiles are nondimensionalized and shown in log-log scales. The vertical axis $z$ is nondimensionalized by significant wave height $H_s$; the horizontal axis $\epsilon$ is nondimensionalized by $H_s$ and horizontally averaged, depth-integrated dissipation $\Psi$, defined as
\begin{equation} \label{eqn:Psi}
\Psi = \langle \int_{-\infty}^{\eta(x,y)} \epsilon dz \rangle =  \langle \int_{-\infty}^{\eta(x,y)} 2\nu s_{ij}s_{ij} dz \rangle.
\end{equation}
This integral has units of $\mathrm{m}^{3}\mathrm{s}^{-3}$, which is the dissipation rate per unit area. Cases of varying steepness collapse well, suggesting that $H_s$ and $\Psi$ are suitable parameters for nondimensionalization and that there may be a universal shape function. We have found that a function 
\begin{equation}
\hat{\epsilon} = 2(|\hat{z}| + 1)^{-3} \label{eqn:shape}
\end{equation} 
describe the data and the shape of the profile well, as shown in Figure \ref{fig:diss-scaling}(b). This is different from the T96 fit in the sense that the near surface dissipation rate is not constant.

We also plot the field measurements from \cite{anis_surface_1995} (AM95), \cite{terray_estimates_1996} (T96), \cite{drennan_oceanic_1996} (D96), \cite{soloviev_observation_2003} (SL03), \cite{sutherland_field_2015,sutherland_measuring_2015} (S15), adapted from a similar Figure in \cite{rascle_note_2013} where the profiles resulting from different breaking entrainment models are discussed in detail. There is a general good agreement between the numerical and observational data in terms of the nondimensionalized profiles, despite the fact that the dissipation rate is computed with very different methods. In addition, in observational studies the depth-integrated dissipation rate $\Psi$ is often estimated using a parameterized wave dissipation (further discussed in Section \ref{sec:dissipation}). The uncertainty in the estimation of $\Psi$ can shift the profiles along the horizontal axis in Figure \ref{fig:diss-scaling}(b), but will not change the general shape of the profile, which is reasonably described by Equation \ref{eqn:shape}. 

Commenting on the shape of the turbulence dissipation rate profile, the functional form eq 14 we propose is compatible near the surface with $\epsilon \propto z^{-1}$. Gradually, the profile transitions to a steeper slope close to $z^{-2}$ below roughly one $H_s$. This empirical power law $z^{-2}$ discussed in various earlier observational studies is also compatible with Equation \ref{eqn:shape} proposed here, and may be interpreted as merely describing the rapid decrease in turbulence intensity below $H_s$. 
In the immediate vicinity of the surface, we confirm the strong but still depth-dependent dissipation, which as commented by S15, does not support the assumption of constant dissipation near the surface made in earlier works (T96).

It is interesting that, despite shear not being the primary production mechanism, but rather wave breaking, the near-surface $\epsilon$ profile exhibits a slope similar to the law-of-the-wall prediction ($z^{-1}$). Due to a lack of wind forcing in the current setup, we cannot directly quantify the enhancement of breaking-induced dissipation compared to a law-of-the-wall baseline, $\epsilon=u_*^3/\kappa z$, as any friction velocity $u_*$ inferred from the wave field would only be nominal. Additionally, strong downward transport of TKE is expected that violates the balance between local shear production and dissipation.

\begin{figure}[h]
    \centering
    \includegraphics[width=1\linewidth]{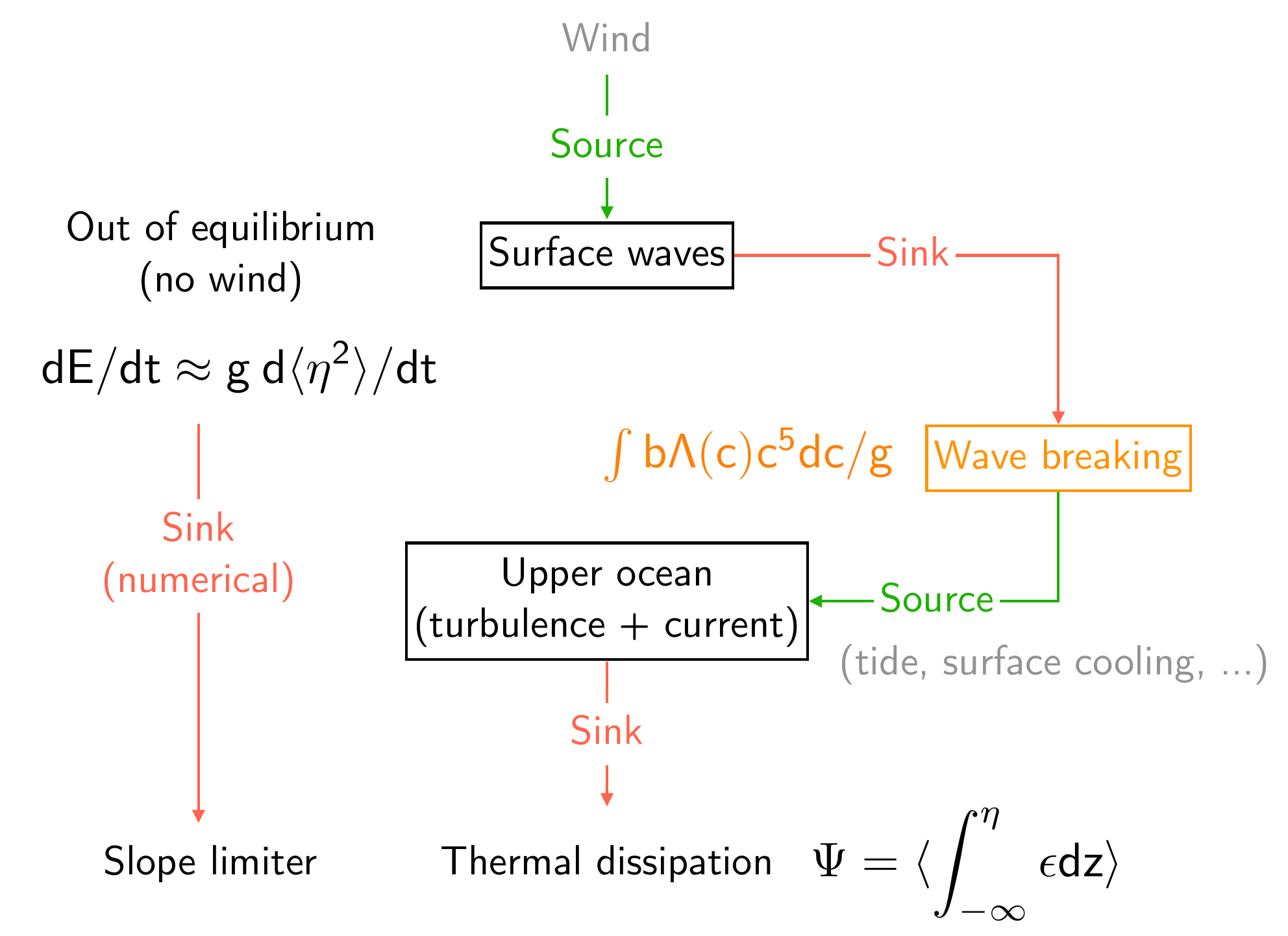}
    \caption{Flow chart of energy pathways. The $dE/dt$ term, the $\Psi$ term, and the fifth moment term all have dimension of dissipation rate per unit area per density ($\mathrm{m^{3}s^{-3}}$).}
    \label{fig:flowchart}
\end{figure}

\section{Energy budget} \label{sec:dissipation}
Finally, we discuss the global energy budget of the system. Figure \ref{fig:flowchart} illustrates the energy pathways between the surface waves and the interior flows (turbulence and current). In reality, the waves gain energy from the wind and dissipate energy through wave breaking. Wave breaking is a source of energy for upper-ocean turbulence (together with other sources), which is eventually dissipated through small-scale thermal dissipation. In the numerical system, there is no wind forcing (in the current setup), and the waves are out of equilibrium. However, the turbulence and current are still in quasi-equilibrium with the source of wave breaking balancing the sink of thermal dissipation. In addition, the numerical slope limiter dissipates a certain amount of energy directly. 

\begin{figure}[h]
    \centering    \includegraphics[width=0.9\linewidth]{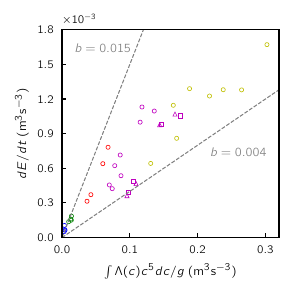}
    \caption{The energy dissipation (measured from global kinetic and potential energy) v.s. the fifth moment of $\Lambda(c)$. The colors corresponds to different cases. Circles: $N_L=15$; squares: $N_L=30$; triangles: $N_L=45$.}
    \label{fig:dEdt_fifth}
\end{figure}


Based on lab-scale experiments and field observations, the dissipation of a broadband wave spectrum can be formulated as \citep{phillips_spectral_1985}
\begin{equation}
S_{ds} (c) = bg^{-1}c^5\Lambda(c)
\end{equation}
That is, the dissipation at a particular wavenumber (related to $c$ by the dispersion relation) is given by the fifth moment of the breaking distribution times the breaking strength parameter $b$. 

We can diagnose a breaking parameter $b$ for the numerical simulations. Figure \ref{fig:dEdt_fifth} shows $dE/dt$ (measured from the energy curves in Figure \ref{fig:energy}(a)), plotted against the fifth moment of $\Lambda(c)$. There is over an order-of-magnitude difference in dissipation rate for cases with different steepness. The corresponding breaking parameter $b$ falls within 0.004 to 0.015, which is comparable to measured values in lab and field. A scale-dependent breaking parameter could also be considered \citep{romero_distribution_2019}.




The (sometimes implicit) assumption that the upper-ocean turbulence is in quasi-equilibrium leads many field studies to estimate $\Psi$ in equation \ref{eqn:Psi} using a spectrally-integrated
(parameterized) dissipation term $S_{ds}$. Some works even make a further assumption that the waves are in equilibrium and estimate \ref{eqn:Psi} using an (also parameterized) wind input source term $S_{in}$. All these implicit assumptions can cause discrepancy for inter-comparison between the simulations and observations, even among observations in different conditions. In the numerical setup, due to the additional dissipative mechanism from the slope limiter, we cannot directly link the global $dE/dt$ to the vertically-integrated thermal dissipation. Therefore, we diagnose $\epsilon$ directly from the velocity gradient computation in the earlier discussion. Because of the nondimensionalization, the shape of the profile in Figure \ref{fig:diss-scaling}(b) should be largely independent from the integrated value of $\Psi$.



\section{Summary and discussion}
We applied a novel multi-layer numerical solver to broad-banded wave breaking and focused on the structure of underwater breaking-induced turbulence and related dissipation. There is a rich field of vorticity, which is a signature of turbulence generation by breaking waves, since the non-breaking wave motion is largely irrotational. 

An advantage of our numerical simulations is that we are able to sample the full 3D velocity field. 
We examined the effect of vertical coordinate choice on derived mean statistics, which has implications for field observations. The surface velocity spectra show strong anisotropy and a mix of wave and turbulence signals. Turbulent dissipation is known to be intermittent and the numerical simulations allow us to compute dissipation rate statistics from spatial averaging over a large area, therefore circumventing the potential sampling uncertainty associated with single-point field observations. 

We diagnosed the vertical profiles of the dissipation rate directly from the strain rate tensor and compared them to field observations. We observe self-similar dissipation rate profiles which have a slope $z^{-1}$ near the surface and a faster than $z^{-1}$ decay below roughly one significant wave height. This confirms that the breaking-induced turbulence is confined to a shallow surface layer and motivates future studies including additional production mechanisms such as wind shear and Langmuir turbulence. We propose a simple empirical functional form to describe the dissipation rate profiles.

Finally, we relate the global energy dissipation to the breaking front statistics, by which we estimate the approximate range of breaking coefficient in our numerical setup using linear regression. 


In the context of ocean surface boundary layer, the layer we describe here is the surface layer, heavily influenced by wind forcing, wave breaking, and wave-turbulence interaction. The current setup was designed to isolate the role of wave breaking in turbulence generation. In reality, there are many mechanisms contributing to the turbulence generation (convection, Langmuir turbulence, near-inertial motion, sub-mesoscale instability, etc.), some of which are able to generate turbulence deeper beyond the surface layer \citep[e.g.,][]{buckingham_contribution_2019, zippel_parsing_2022}. Our study is idealized and limited because it does not include all these other mechanisms for turbulence generation in the surface layer. However, the idealized setup here is a useful tool to examine the role of wave breaking, allowing us to distinguish the physics of the free-surface boundary layer from the wall layer, precisely because it does not include all these other mechanisms for turbulence generation in the surface layer. 


\acknowledgments
This work is supported by the \textit{National Science Foundation} under grant 2318816 to LD (Physical Oceanography program), the \textit{NASA Wind Vector Science Team}, grant 80NSSC23K0983 to LD and JTF. We thank Peter Sutherland for providing the data and scripts for the turbulence dissipation rate data used in Figure \ref{fig:diss-scaling} and originally shown in \cite{sutherland_field_2015}.
%
%
\datastatement
The Basilisk solver is open source (\url{http://basilisk.fr/}), and the specific setup and post-processing code related to this work can be found at \url{https://github.com/jiarong-wu/multilayer_breaking}.






%
%
%


\bibliographystyle{ametsoc2014}
\bibliography{ref}

\end{document}